\documentclass[conference]{IEEEtran}

\usepackage[T1]{fontenc}
\usepackage[utf8]{inputenc}
\usepackage{newtxtext,newtxmath}
\usepackage{cite}
\usepackage{amsmath}
\usepackage{graphicx}
\usepackage{float}
\usepackage{booktabs}
\usepackage{tabularx}
\usepackage{array}
\usepackage{xcolor}
\usepackage{url}
\usepackage{balance}
\usepackage{placeins}
\usepackage{microtype}
\usepackage{hyperref}
\usepackage{pgfplots}
\pgfplotsset{compat=1.18}
\hypersetup{colorlinks=true,linkcolor=black,citecolor=black,urlcolor=black}

\definecolor{jevblue}{RGB}{0,114,178}
\definecolor{geminiorange}{RGB}{230,159,0}
\definecolor{nvidiagreen}{RGB}{0,158,115}
\definecolor{metapink}{RGB}{204,121,167}


\newcolumntype{Y}{>{\raggedright\arraybackslash}X}

\title{JEV vs. LLMs: A Comparative Analysis for Phishing Email Detection}

\author{
\parbox[t]{0.49\textwidth}{\centering
Muhammad Mahir Malik\textsuperscript{*}\\
\mbox{\textit{Department of Software Engineering}}\\
\mbox{\textit{University of Central Punjab}}\\
Lahore, Pakistan\\
L1F23BSSE0016@ucp.edu.pk}
\hfill
\parbox[t]{0.49\textwidth}{\centering
Muhammad Moiz Khalid\\
\mbox{\textit{Department of Artificial Intelligence and Data Science}}\\
\mbox{\textit{National University of Computer and Emerging Sciences}}\\
Islamabad, Pakistan\\
i232552@isb.nu.edu.pk}
\\[2ex]
\parbox[t]{0.49\textwidth}{\centering
Hussain Ahmad\\
\mbox{\textit{School of Computer Science and Information Technology}}\\
\mbox{\textit{Adelaide University}}\\
Adelaide, Australia\\
hussain.ahmad@adelaide.edu.au}
\hfill
\parbox[t]{0.49\textwidth}{\centering
Faheem Ullah\\
\mbox{\textit{College of Interdisciplinary Studies}}\\
\mbox{\textit{Zayed University}}\\
Abu Dhabi, UAE\\
faheem.ullah@zu.ac.ae}

\thanks{* Corresponding author.}
}
\IEEEoverridecommandlockouts
\begin{document}

\pagestyle{plain}
\thispagestyle{plain}

\maketitle

\begin{table*}[!t]
\caption{Comparison of Existing Studies with the Proposed Study}
\label{tab:related}
\centering
\small
\setlength{\tabcolsep}{4pt}
\begin{tabularx}{\textwidth}{@{}p{1.55in}p{0.4in}p{1.75in}p{1.6in}Y@{}}
\toprule
\textbf{Study} & \textbf{Year} & \textbf{Approach} & \textbf{Task or data} & \textbf{Gap relative to this study} \\
\midrule
Singh et al.~\cite{singh} & 2022 & Deep learning & Phishing emails & Proposes a new detector \\
Atawneh and Aljehani~\cite{atawneh} & 2023 & Deep learning & Phishing emails & Proposes a new detector \\
Thakur et al.~\cite{thakur} & 2023 & Systematic review & Phishing email literature & No matched operational comparison \\
Patel et al.~\cite{patel} & 2024 & 15 prompted LLMs & Scam and phishing emails & No cost, latency, or calibration \\
Lee~\cite{lee} & 2025 & Prompt-based LLM & Phishing emails & LLM only; no operational comparison \\
Ji and Kim~\cite{ji} & 2025 & LLM-based detection models & Phishing detection & No non-LLM decision model \\
Nguyen et al.~\cite{nguyen} & 2025 & Debate-driven multi-agent LLMs & Phishing emails & LLM only \\
Murthy et al.~\cite{murthy} & 2025 & ML, ANN, and boosting & 11,430 phishing URLs & URL input rather than email \\
Kuikel et al.~\cite{kuikel} & 2025 & Fine-tuned LLMs & Detection and explanation & No structured non-LLM model \\
Uddin et al.~\cite{uddin} & 2026 & Explainable transformer & Phishing emails & Proposes a new model \\
Bryo AI remark~\cite{infoq} & 2026 & Informal Jev and Gemini comparison & Email classification & Informal; limited metric coverage \\
\textbf{This study} & 2026 & Jev 1.13 and three LLMs & 1500 labelled emails & Five evaluation dimensions together \\
\bottomrule
\end{tabularx}
\end{table*}

\begin{abstract}
Phishing remains a major source of credential theft and fraudulent payments, and email security systems must classify large message volumes reliably, quickly, and at manageable cost. Large Language Models (LLMs) offer flexible text understanding, but their token-based generation increases response time and API cost, and the reliability of their confidence values requires measurement. Jev 1.13, a typed decision model from TypeSafe AI, instead returns a structured outcome with a confidence value. This paper compares Jev 1.13 with Gemini 3.5 Flash-Lite, NVIDIA Nemotron 3 Ultra, and Meta Llama 4 Maverick on a labelled dataset of 1500 emails, covering classification performance, decision latency, token consumption, API cost, and probability calibration. Jev achieves the highest accuracy (79.67\%), precision (71.43\%), recall (97.97\%), and F1 score (82.62\%). It also records the shortest batch time (173.88 seconds), the lowest token count (597,895), the lowest cost (USD 0.015103 per 1000 decisions), and the lowest Brier score (0.1698) and ECE (0.1594). These results suggest that Jev is a promising option for fast, structured, high-volume triage. LLMs retain the advantage of generating explanations, which we do not evaluate, and all four systems require further validation before standalone deployment.
\end{abstract}

\begin{IEEEkeywords}
phishing detection, email security, large language models, Jev, classification performance, decision latency, API cost, probability calibration
\end{IEEEkeywords}

\section{Introduction}\label{sec:intro}

Generative AI, particularly LLMs, has gained significant attention across various domains \cite{abbas2025scalar, zhang2026mrif}, including cybersecurity \cite{zhang2026explainable, goel2025co}, cloud computing \cite{ahmad2025resilient, ahmad2024smart}, and financial systems \cite{zhang2025regimefolio, chen20253s}. In cybersecurity, LLMs are increasingly being explored for vulnerability assessment \cite{jayalath2024microservice, arifin2026agenticvm}, situational awareness \cite{ahmad2025survey, ahmad2023review}, and risk management \cite{chopra2026chatnvd, abdulsatar2025towards}. Despite these advancements, phishing emails remain a pervasive cybersecurity threat \cite{goel2024machine}, enabling attackers to steal sensitive information, compromise credentials, conduct financial fraud, and distribute malicious links. The Anti-Phishing Working Group (APWG) records 1,003,924 phishing attacks in the first quarter of 2025, its highest quarterly total since late 2023. The report also identifies online payment and financial/banking services as major targets and reports a 33\% quarterly increase in wire-transfer Business Email Compromise (BEC) attacks~\cite{apwg}. The FBI Internet Crime Complaint Center's 2025 report lists 1,008,597 cybercrime complaints and USD 20.877 billion in reported losses. It also records 191,561 phishing and spoofing complaints and nearly USD 3.05 billion in losses associated with BEC~\cite{ic3}. These figures establish the scale and financial impact of email-based threats.

Deploying phishing detectors in operational settings raises requirements that extend beyond classification accuracy. A security team may process thousands of incoming emails and must reach timely decisions without overwhelming analysts with false alarms. A detector with high recall but poor precision sends too many legitimate messages to manual review, whereas a detector with strong classification performance but high latency or cost becomes difficult to operate at scale. Earlier detectors apply conventional machine-learning classifiers to features derived from URLs~\cite{murthy}, whereas deep-learning approaches learn phishing patterns directly from email content~\cite{atawneh,singh,uddin,thakur}. LLM-based approaches add flexible text understanding and can generate explanations~\cite{ji,nguyen,patel,lee,kuikel}, but the reliability of their outputs and the calibration of their confidence values require measurement~\cite{li,darwish,detommaso,guo,lane}, as do their response time and cost at scale. Jev 1.13 offers a different interface: TypeSafe AI describes it as a typed decision model that returns structured outcomes and confidence values rather than a free-form text response~\cite{vktr,infoq,typesafe}. This design motivates a direct comparison, but it does not by itself demonstrate superior performance.

In this paper, we present an empirical comparison of Jev 1.13 against three LLMs, namely Gemini 3.5 Flash-Lite, NVIDIA Nemotron 3 Ultra, and Meta Llama 4 Maverick. All four systems receive the same 1500-email Kaggle dataset, in the same order, under a common binary classification task. We do not propose a new detector. Instead, we evaluate the systems on classification performance, decision latency, token consumption, API cost, and probability calibration, so that decision quality can be read together with operational behaviour. The study addresses the following three research questions:

\textbf{RQ1:} How does the phishing classification performance of Jev 1.13 compare with LLMs?

\textbf{RQ2:} How do Jev 1.13 and LLMs compare in terms of computational efficiency and cost-effectiveness in phishing email classification?

\textbf{RQ3:} How closely do their reported confidence values align with classification correctness, as measured by the Brier score and expected calibration error (ECE)?

The main contributions of this paper are summarized as follows:
\begin{itemize}
  \item Designing a reproducible comparison workflow that applies the same labelled email set, in the same order, to Jev 1.13 and three LLMs under a common binary classification task.
  \item Implementing the workflow through API-based Python scripts that record the predicted class, reported confidence, response time, and token counts for every decision.
  \item Evaluating the four systems on classification, latency, token, cost, and calibration measures, and reporting the results in comparable evaluation tables.
  \item Interpreting the results as evidence about the suitability of structured decision models and LLMs for different parts of a phishing triage pipeline, while identifying the limits of the current experiment.
\end{itemize}

The rest of this paper is organized as follows. Section~\ref{sec:lit} reviews related work on phishing detection, LLM reliability, and Jev, and identifies the evaluation gap addressed here. Section~\ref{sec:method} describes the research methodology, including the model configurations, the inference workflow, and the evaluation metrics. Section~\ref{sec:setup} presents the dataset and the experimental setup, and Section~\ref{sec:results} reports the results. Section~\ref{sec:discussion} discusses the findings, Section~\ref{sec:threats} examines threats to validity, and Section~\ref{sec:conclusion} concludes the study and outlines future work.

\section{Literature Review}\label{sec:lit}

Phishing email detection has been studied through traditional machine learning, deep learning, and, more recently, generative language models. These studies differ in input type, model family, and evaluation measures, which limits how directly their results can be compared. In this section, we group the relevant work by approach and relate each group to the evaluation reported in this paper.

Table~\ref{tab:related} summarizes the reviewed studies and clarifies the gap that our evaluation addresses. Previous work contributes important pieces of the phishing-detection problem, but it uses different input types, model families, and evaluation measures. URL-based detection, email classification, and explanation assessment answer related but distinct questions. Compared with these studies, the present work makes no architectural proposal. It applies one labelled email set to Jev 1.13 and three LLMs and reports classification, latency, token, cost, and calibration results together. We therefore position our contribution as a common-condition comparison of decision quality and operational behaviour, rather than as a new detection architecture.

\begin{figure}[!t]
\centering
\includegraphics[width=\columnwidth]{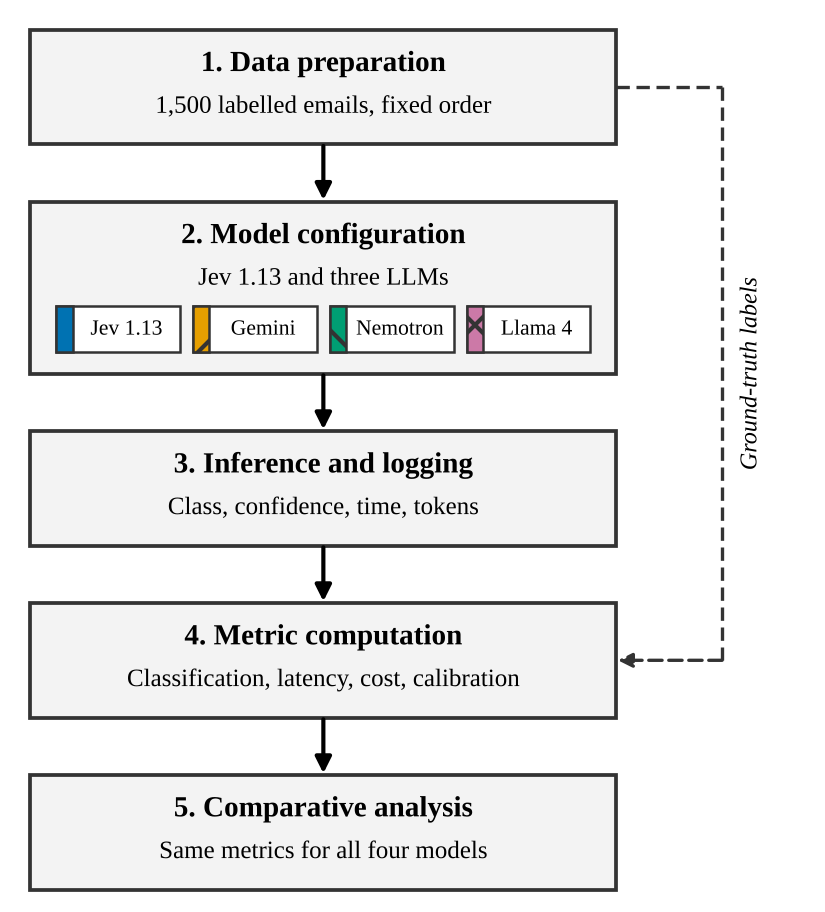}
\caption{Research Methodology.}
\label{fig:workflow}
\end{figure}

\subsection{Deep-Learning and Classical Detectors}
Atawneh and Aljehani~\cite{atawneh} and Singh et al.~\cite{singh} present deep-learning models that learn phishing patterns from email content. Both works concentrate on building a detection model. In contrast, our study does not train a new detector and instead compares existing systems under shared conditions.

Thakur et al.~\cite{thakur} review deep-learning phishing email detectors and identify recurring concerns involving dataset size, label consistency, language coverage, and adaptation to new attack patterns. These concerns show why classification performance must be interpreted in the context of dataset design and evaluation scope. However, the review does not provide a matched latency, cost, and calibration comparison of the systems studied here.

Murthy et al.~\cite{murthy} compare classical classifiers, artificial neural networks, and boosting methods on 11,430 phishing URLs and report advantages for ensemble approaches. Their URL-based input differs from full email-message classification, which limits direct comparison with our setting. Uddin et al.~\cite{uddin} present an explainable transformer-based model for phishing email detection built on a large language model approach. Their emphasis is on model design and explainability, whereas our study does not propose a model and instead adds operational and calibration measures.

\subsection{LLM-Based Phishing Detection}
Research on LLMs for phishing explores direct prompting, fine-tuning, and explanation generation. Patel et al.~\cite{patel} evaluate 15 LLMs on scam and phishing email identification, and Ji and Kim~\cite{ji} examine the effectiveness of LLM-based phishing detection models. Both studies examine how well LLMs detect phishing, while our comparison adds a non-LLM typed decision model together with token, API cost, latency, and calibration measures.

Lee~\cite{lee} examines prompt-based LLM classification of phishing emails and discusses accuracy and the value of human-readable explanations, while Nguyen et al.~\cite{nguyen} study debate-driven multi-agent LLMs for phishing email detection. Both studies are centred on LLMs, whereas our comparison places a structured decision model alongside three LLMs and reports operational measures.

Kuikel et al.~\cite{kuikel} evaluate fine-tuned LLMs using accuracy alongside self-consistency, faithfulness, and explainability, and they highlight that a plausible explanation does not necessarily guarantee a correct classification. This motivates an evaluation that distinguishes the final label from any explanation a model generates. However, the study does not include a structured non-LLM decision model in the comparison.

\subsection{Reliability and Calibration of LLM Outputs}
Operational reliability also matters. Li et al.~\cite{li} survey hallucination in LLMs and describe the risk of unsupported generated content, and Darwish et al.~\cite{darwish} examine a multi-agent framework for mitigating hallucinations. Detommaso et al.~\cite{detommaso} study confidence calibration for LLMs, Guo et al.~\cite{guo} provide foundational work on the calibration of modern neural networks, and Lane~\cite{lane} reviews classifier probability-calibration measures. These works are not specific to phishing, but together they support measuring not only whether a model predicts the correct class but also whether its reported confidence is informative.

\subsection{Jev and the Evaluation Gap}
TypeSafe AI presents Jev 1.13 as a model that returns typed probabilistic decisions instead of generated prose~\cite{vktr,typesafe}. Independent reporting discusses its speed and cost characteristics, while also noting that measured gains depend on workload and evaluation conditions~\cite{infoq}. The same report relays an informal practitioner comparison, attributed to Bryo AI, that describes a cost advantage for Jev over Gemini but does not provide the broad metric coverage used in this study~\cite{infoq}. We therefore treat these reports as motivation and not as a substitute for a controlled comparison.

\section{Research Methodology}\label{sec:method}

\subsection{Research Design and Scope}

In this study, we adopt a comparative experimental methodology to systematically evaluate Jev 1.13 against three LLMs for phishing email classification. The objective is to investigate their relative performance, operational efficiency, and associated trade-offs. To ensure a fair and consistent comparison, all four systems are evaluated using the same labelled dataset and a common set of evaluation metrics, with system choice serving as the primary comparison variable. We formulate phishing email detection as a binary classification problem, where phishing and legitimate emails represent the positive and negative classes, respectively.

The research methodology is structured into five sequential stages: (i) dataset acquisition and label preparation, (ii) model configuration, (iii) inference and output logging, (iv) evaluation metric computation, and (v) comparative analysis. These stages establish a systematic evaluation framework that ensures experimental consistency, transparency, and reproducibility. The following subsections describe each stage in detail.



\begin{table}[!t]
\caption{Evaluation Dimensions and Metrics}
\label{tab:metrics}
\centering
\small
\begin{tabularx}{\columnwidth}{@{}p{0.85in}p{1.1in}Y@{}}
\toprule
\textbf{Dimension} & \textbf{Measures} & \textbf{Purpose} \\
\midrule
Classification & Accuracy, precision, recall, F1 & Correctness and error types \\
Latency & Batch, first-response, \& avg. response time & Speed at scale \\
Tokens & Input, output, and total tokens & API usage \\
Cost & Actual cost, cost per 1000 decisions & Financial burden \\
Calibration & Brier score, ECE & Reliability of confidence \\
\bottomrule
\end{tabularx}
\end{table}

\subsection{System Components and Data Flow}
Figure~\ref{fig:workflow} outlines how the five stages connect. Conceptually, the workflow can be described in terms of four components. A data component holds the labelled emails and supplies them to every model in the same fixed order. A model-interface component sends each email to one of the four systems: for Jev, through the TypeSafe AI API under a Choice-type decision schema, and for the three LLMs, through the respective provider APIs with the shared classification prompt. A logging component records the predicted class, the reported confidence or probability, the response time, and the input and output token counts for every decision. An analysis component then reads these logs, aligns each prediction with its ground-truth label, and computes the classification, latency, token, cost, and calibration measures reported in Section~\ref{sec:results}. Data therefore flow in one direction, from the labelled corpus through the model interfaces and logs to the metrics, and no result is fed back into any model.

\begin{figure*}[!t]
\centering
\includegraphics[width=\textwidth]{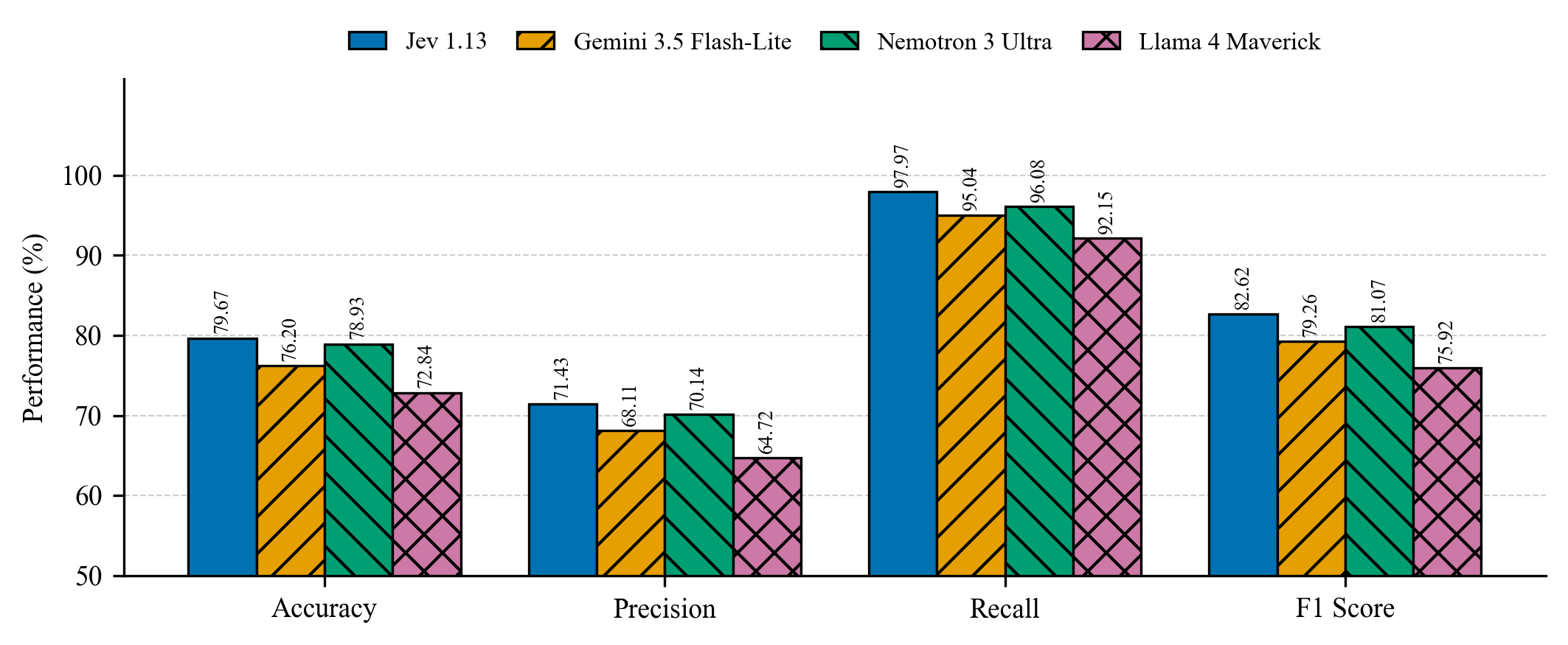}
\caption{Comparison of Jev and LLMs in phishing classification.}
\label{fig:classification}
\end{figure*}

\subsection{Model Configuration}
We compare four model configurations. Jev 1.13 (TypeSafe AI) is configured through the TypeSafe AI API with a Choice-type decision schema, in which each email is submitted as the state and the model returns one of two predefined class labels together with a confidence value~\cite{typesafe}. Gemini 3.5 Flash-Lite (Google), NVIDIA Nemotron 3 Ultra (NVIDIA), and Meta Llama 4 Maverick (Meta) are each accessed through chat-completion APIs, with the same email fields and the same binary classification task submitted to every model. The NVIDIA model is called through the OpenRouter endpoint with the identifier \texttt{nvidia/nemotron-3-ultra-550b-a55b:free}, and the LLM scripts set the temperature to 0 so that outputs are as deterministic as the providers allow.

We keep the classification objective and email input consistent across the four configurations. For the three LLMs, we use one equivalent prompt template rather than model-specific prompt optimization. A system message instructs the model to act as an email phishing classification system and to follow the requested output format. The user message asks the model to classify the email as phishing or legitimate using only the supplied information, lists the input fields described in Section~\ref{sec:setup}, and requires the following output format:
\begin{center}
\begin{minipage}{0.85\columnwidth}
\footnotesize
\begin{verbatim}
LABEL: phishing
CONFIDENCE: 0.95
REASON: short explanation
\end{verbatim}
\end{minipage}
\end{center}
The label must be either phishing or legitimate, and the confidence must be a number between 0 and 1 that expresses confidence in the selected label. This choice supports a controlled baseline comparison, although it does not establish each LLM's best achievable performance. Jev returns a typed decision, whereas the LLMs return generated responses that we interpret as binary decisions with confidence values.

\subsection{Output Parsing and Probability Mapping}
For the LLMs, we convert each generated response into a binary decision and a phishing probability. The response is lower-cased, and the label is read from the \texttt{LABEL} line. If that line is absent, the first occurrence of the word phishing or legitimate in the text is used, and a response containing neither is recorded as unknown. The confidence is read from the \texttt{CONFIDENCE} line and defaults to 0.5 if it cannot be parsed. Because the reported confidence refers to the selected label, the phishing probability $p$ equals the confidence for a phishing prediction and one minus the confidence for a legitimate prediction, and it equals 0.5 for an unknown prediction. Jev returns its label and confidence directly through the typed schema, and we convert them in the same way. Unknown predictions and requests that still fail after retries are excluded from the scored decisions.

\subsection{Inference Workflow and Recorded Outputs}
We run Python scripts that call the TypeSafe AI API for Jev and the respective provider APIs for the three LLMs, and we submit the 1500 emails to each model under the same evaluation conditions. For each decision, we record the predicted class, the reported confidence or probability, the response time, the input and output token counts, and the actual cost reported in the API response. These records support classification scoring, latency calculation, token accounting, API-cost calculation, and calibration analysis.

We process the logged outputs in five steps. First, we align each model's predicted class with the same ground-truth label. Second, we calculate accuracy, precision, recall, and F1 score. Third, we aggregate response times to obtain total batch time, first-response time, and average individual response time. Fourth, we sum input and output tokens and normalize the actual API cost to 1000 decisions. Fifth, we compute the Brier score and expected calibration error from the recorded phishing probabilities. We then consolidate the outputs into the comparative tables and figures in Section~\ref{sec:results}.

\subsection{Evaluation Dimensions and Metrics}
We group the experiment into five evaluation dimensions, which Table~\ref{tab:metrics} summarizes together with the reason each one matters for an operational phishing filter. We use complementary measures because a model can achieve strong overall accuracy while still missing phishing messages, flagging legitimate messages, responding slowly, or reporting poorly calibrated confidence. Token counts serve as a supporting resource measure because they help explain API consumption and cost.

For classification, we use the standard definitions of accuracy, precision, recall, and F1 score with phishing as the positive class. Accuracy summarizes overall correctness, precision reflects the reliability of phishing flags, recall measures how many phishing emails the system identifies, and F1 balances precision and recall. For latency, each request is timed from submission to response, the total batch time covers the full corpus, and the first-response time is the time of the first request. We report the average individual response time over the scored decisions.

The Brier score is the mean squared difference between the predicted phishing probability $p_i$ and the binary ground-truth label $y_i$ over $N$ scored decisions:
\begin{equation}
\mathrm{BS}=\frac{1}{N}\sum_{i=1}^{N}(p_i-y_i)^2.
\end{equation}
ECE groups the decisions into $M=10$ equal-width probability bins $B_m$ and weights the gap between the observed phishing rate $\bar{y}(B_m)$ and the mean predicted probability $\bar{p}(B_m)$ in each bin:
\begin{equation}
\mathrm{ECE}=\sum_{m=1}^{M}\frac{|B_m|}{N}
\left|\bar{y}(B_m)-\bar{p}(B_m)\right|.
\end{equation}
We normalize the actual cost of the run to 1000 decisions as
\begin{equation}
C_{1000}=\frac{C_{\mathrm{total}}}{N}\times 1000,
\end{equation}
where $N=1{,}500$ in our runs. Lower values of the Brier score and ECE indicate better calibration.

\begin{figure*}[!t]
\centering

\begin{minipage}[t]{0.48\textwidth}
\centering
\vspace{0pt}
\includegraphics[width=\linewidth]{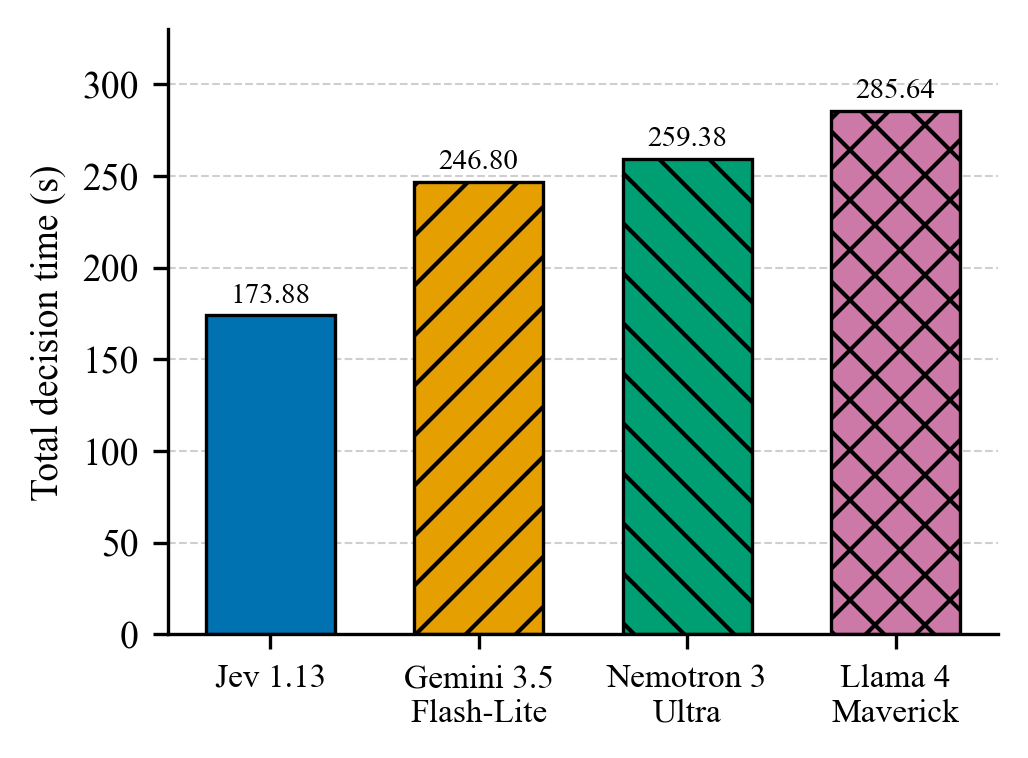}
\caption{Comparison of total decision time between Jev and LLMs.}
\label{fig:latency}
\end{minipage}
\hfill
\begin{minipage}[t]{0.48\textwidth}
\centering
\vspace{0pt}
\includegraphics[width=\linewidth]{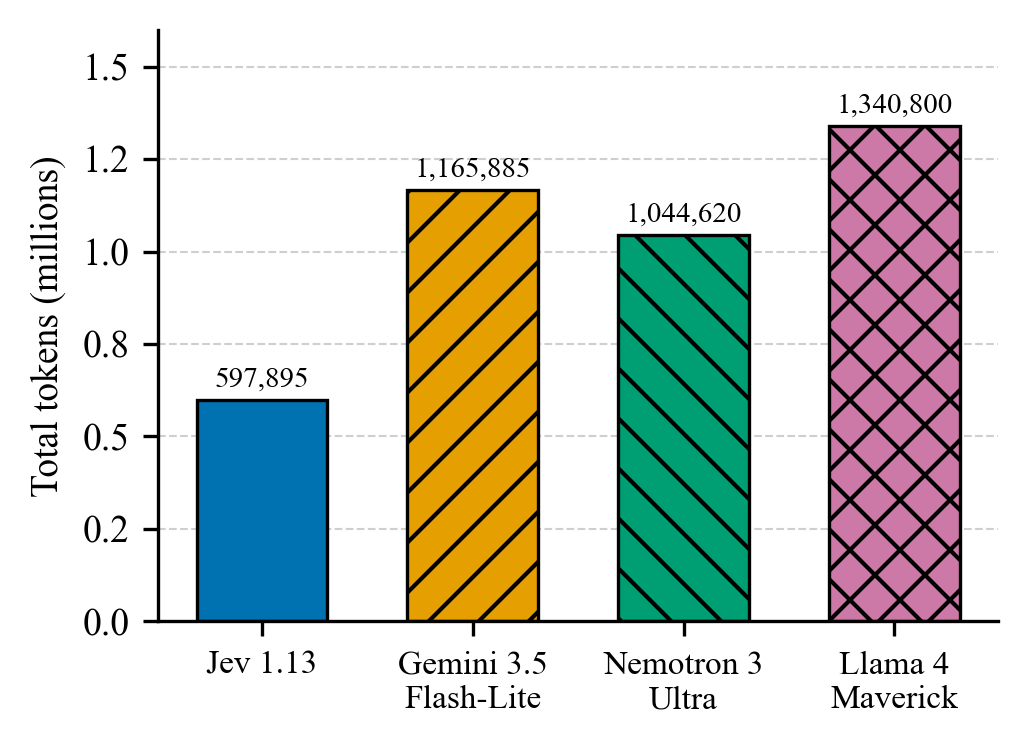}
\caption{Comparison of total token consumption between Jev and LLMs.}
\label{fig:tokens}
\end{minipage}

\end{figure*}

\section{Experimental Setup}\label{sec:setup}

\subsection{Dataset Selection and Preparation}
We select the \textit{Phishing Email Detection 2026} dataset because its labelled email records directly match our binary research task of distinguishing phishing messages from legitimate messages. The dataset is obtained from Kaggle, and it provides the class labels needed to score every model against a common reference~\cite{kaggle}. We choose this resource for its task alignment, public listing, labelled examples, and suitability for a controlled same-input comparison. A fixed corpus also lets us examine several complementary properties of one decision process, namely correct classification, false-alarm behaviour, missed phishing messages, response time, cost, and confidence quality. We do not describe the dataset as a widely adopted benchmark.

Each record provides an email identifier, the sender address, the subject line, whether the email contains a link, whether it contains an attachment, an urgency score, a spelling-error count, the email length in words, and a binary phishing label. The models receive these fields, and the label serves only as the ground-truth reference. A label value of 1, true, yes, or phishing is mapped to the phishing class, and any other value is mapped to legitimate. We inspect the records and verify that all required columns are present before evaluation. The dataset contains 1500 emails, and Table~\ref{tab:dataset} shows its class distribution. We use all records in their original order for all four systems, with no separate training and test split, because no model is trained and all four systems are used as provided through their APIs.

\begin{table}[!t]
\caption{Dataset Class Distribution}
\label{tab:dataset}
\centering
\small
\begin{tabular}{lrr}
\toprule
Class & Count & Percentage (\%) \\
\midrule
Phishing & 740 & 49.33 \\
Legitimate & 760 & 50.67 \\
Total & 1500 & 100.00 \\
\bottomrule
\end{tabular}
\end{table}

\subsection{Environment and Execution}
We execute the experiment locally from a Visual Studio Code terminal using Python scripts. The scripts connect directly to the TypeSafe AI API for Jev 1.13 and to the respective provider APIs for Gemini 3.5 Flash-Lite, NVIDIA Nemotron 3 Ultra, and Meta Llama 4 Maverick, and we authenticate requests with API keys. We use a standard internet connection rather than a dedicated low-latency environment, so the measured time includes the practical effects of API communication and network conditions. Each request has a timeout of 120 seconds. A request that returns a rate-limit or server error is retried up to three times with exponential backoff, and a request that still fails is logged as an error. The scripts save interim results every 50 emails and write the per-email predictions, confidences, token counts, costs, and response times to a results file.

\subsection{Comparison Conditions and Evaluation Focus}
Every model receives the same email fields in the same order, and each model solves the same phishing-versus-legitimate task. We run one evaluation pass per model. We use the same prompt template across the LLMs to maintain a common baseline and configure Jev through its typed Choice schema. The three LLMs act as the comparison baselines for Jev, and no additional baselines, such as classical classifiers, are included.

We measure classification performance, latency, token consumption, API cost, and probability calibration. We report total time for the full batch, first-response time, and average individual response time, and we calculate the cost per 1000 decisions from the actual cost of the 1500-email run. Table~\ref{tab:metrics} gives the rationale for each measure.

\section{Experimental Results}\label{sec:results}

\subsection{Classification Performance}
Figure~\ref{fig:classification} reports the classification results. Jev records the highest value for all four measures, with 79.67\% accuracy, 71.43\% precision, 97.97\% recall, and 82.62\% F1. NVIDIA Nemotron 3 Ultra follows Jev most closely on accuracy and F1, with 78.93\% and 81.07\%, respectively. Gemini records 76.20\% accuracy and 79.26\% F1, while Meta Llama 4 Maverick records 72.84\% accuracy and 75.92\% F1.

Figure~\ref{fig:classification} shows that all four systems recover most phishing emails but differ in how many legitimate emails they flag. Recall lies between 92.15\% and 97.97\%, whereas precision lies between 64.72\% and 71.43\%. Because precision is lower than recall for every system, each one produces more false positives than false negatives. Jev's 97.97\% recall is the strongest result in this run and indicates fewer missed phishing messages than the other systems. Its 71.43\% precision, however, shows that a meaningful share of the emails it flags as phishing are legitimate. We therefore interpret the result as support for triage rather than for automatic blocking without safeguards.

Relative to the mean of the three LLMs, Jev leads by approximately 3.7 percentage points in accuracy, 3.8 points in precision, 3.5 points in recall, and 3.9 points in F1. The recall bars are high for all four systems, and the lower precision bars explain why a detector of this kind still needs a review or escalation policy.

\begin{table}[!b]
\caption{Operational Efficiency: Latency, Token Usage, and API Cost}
\label{tab:operations}
\centering
\scriptsize
\setlength{\tabcolsep}{1.8pt}
\renewcommand{\arraystretch}{1.08}
\begin{tabular}{@{}p{0.91in}rrrr@{}}
\toprule
\textbf{Measure} & \shortstack{\textbf{Jev}\\\textbf{1.13}} & \shortstack{\textbf{Gemini 3.5}\\\textbf{Flash-Lite}} & \shortstack{\textbf{NVIDIA}\\\textbf{Nemotron 3}} & \shortstack{\textbf{Meta}\\\textbf{Llama 4}} \\
\midrule
Total batch time (s) & 173.88 & 246.80 & 259.38 & 285.64 \\
Total batch time (min) & 2.90 & 4.11 & 4.32 & 4.76 \\
First-response time (s) & 2.26 & 3.92 & 3.18 & 3.74 \\
Avg. response time (s) & 0.58 & 1.25 & 1.04 & 1.31 \\
Input tokens & 539,395 & 1,024,660 & 912,340 & 1,184,320 \\
Output tokens & 58,500 & 141,225 & 132,280 & 156,480 \\
Total tokens & 597,895 & 1,165,885 & 1,044,620 & 1,340,800 \\
Actual cost (USD) & 0.02265459 & 0.04872130 & 0.13892000 & 0.15240000 \\
Cost per 1000 (USD) & 0.015103 & 0.032481 & 0.092613 & 0.101600 \\
\bottomrule
\end{tabular}
\end{table}

\subsection{Decision Latency, Token Consumption, and Cost}
Table~\ref{tab:operations} combines the operational measures. Jev processes the full dataset in 173.88 seconds, compared with 246.80 seconds for Gemini, 259.38 seconds for NVIDIA Nemotron 3 Ultra, and 285.64 seconds for Meta Llama 4 Maverick. The average LLM batch time is 263.94 seconds, so Jev's total time is approximately 34.1\% lower. Jev also records the lowest first-response time (2.26 seconds) and average individual response time (0.58 seconds).

The operational table separates three related but non-identical questions. Batch time measures how long the complete corpus takes, first-response time indicates when the first decision becomes available, and average individual response time summarizes per-email response duration. The token and cost rows add usage and financial dimensions. Jev leads on each reported operational measure, although the measured latency includes network and provider-service effects rather than isolated model inference time.

Jev's actual API cost for the full dataset is USD 0.02265459, equivalent to USD 0.015103 per 1000 decisions. The mean LLM cost per 1000 decisions is USD 0.075565, so Jev's normalized cost is approximately 80.0\% lower than the LLM average, or about one-fifth of it. The recorded Jev cost corresponds to its 539,395 input tokens priced at USD 0.042 per million tokens, the input rate reported for Jev with no charge for output tokens~\cite{infoq}, which explains why its cost is low relative to its token total. NVIDIA Nemotron 3 Ultra records the highest classification scores among the LLMs, yet its normalized cost (USD 0.092613) is higher than Gemini's (USD 0.032481). Accuracy alone therefore does not describe the operational trade-off.

Figure~\ref{fig:latency} isolates the batch-time comparison from the token and cost measures. Jev has the shortest bar at 173.88 seconds, while Meta Llama 4 Maverick has the longest at 285.64 seconds. The ordering agrees with the latency values in Table~\ref{tab:operations}. Because all calls use external APIs over a standard internet connection, the figure reflects the complete request-response path recorded in the experiment, and it should be read as a result from this recorded run rather than as a universal speed guarantee.

Figure~\ref{fig:tokens} focuses on reported token totals rather than elapsed time. Jev records 597,895 tokens, compared with 1,044,620 for NVIDIA Nemotron 3 Ultra, 1,165,885 for Gemini, and 1,340,800 for Meta Llama 4 Maverick. Jev therefore uses approximately 42.8\% fewer tokens than the lowest-token LLM in this experiment (NVIDIA Nemotron 3 Ultra) and approximately 48.7\% fewer than Gemini. We report the totals as recorded by the API calls and interpret them cautiously, because different providers may tokenize and account for requests differently. They are not a universal measure of computational work across model providers.

\begin{table}[!t]
\caption{Probability Calibration Results}
\label{tab:calibration}
\centering
\small
\setlength{\tabcolsep}{3pt}
\begin{tabular}{@{}lrrrr@{}}
\toprule
\textbf{Metric} & \textbf{Jev} & \textbf{Gemini} & \textbf{NVIDIA} & \textbf{Meta} \\
\midrule
Brier score (lower is better) & 0.1698 & 0.2013 & 0.1762 & 0.2318 \\
ECE (lower is better) & 0.1594 & 0.1827 & 0.1618 & 0.2176 \\
\bottomrule
\end{tabular}
\end{table}

\subsection{Probability Calibration}
Table~\ref{tab:calibration} compares the calibration measures. Jev records the lowest Brier score (0.1698) and ECE (0.1594). NVIDIA Nemotron 3 Ultra follows with 0.1762 and 0.1618, respectively. Gemini records 0.2013 and 0.1827, while Meta Llama 4 Maverick records 0.2318 and 0.2176. Since lower values indicate better calibration, the reported confidence values from Jev show the closest alignment with observed correctness among the four systems in this run.

Calibration matters because a security workflow may use confidence to decide whether to accept a prediction automatically or send it for review. Jev has the smallest error under both measures, and NVIDIA Nemotron 3 Ultra ranks second. The gap between Jev and NVIDIA Nemotron 3 Ultra is small on ECE (0.1594 against 0.1618), so repeated experiments are necessary before treating this ordering as stable.


In summary, Jev leads in the reported classification measures and also records the shortest total batch time, the lowest total token count, the lowest normalized API cost, and the lowest calibration errors. NVIDIA Nemotron 3 Ultra is the closest LLM in accuracy, F1, and calibration, whereas Gemini is the least costly of the three LLMs.

\section{Discussion}\label{sec:discussion}
\subsection{What the Results Indicate}
In response to RQ1, our results indicate that a typed decision model can perform strongly on a narrowly defined binary classification task without generating a natural-language explanation for every email. Jev's advantage appears across several dimensions in this run rather than depending on a single classification score. Its recall of 97.97\% is useful for reducing missed phishing messages, while its 71.43\% precision still leaves a role for review, policy thresholds, or additional checks before a message is blocked automatically.

In response to RQ2, the latency and cost results matter for high-volume workflows. Jev completes the 1500-email run in 173.88 seconds and costs USD 0.015103 per 1000 decisions. The three LLMs provide more natural-language output, but their reported batch times and costs are higher in this experiment. The token totals offer another view of this difference, although token counts across providers do not represent identical computational operations. 

In response to RQ3, Jev's lower Brier score and ECE suggest that its confidence values are the most informative of the four in this run, which is relevant when confidence is used to route messages between automatic handling and manual review. The margin over NVIDIA Nemotron 3 Ultra is small, however, and should be read with caution.

\subsection{Relation to Prior Reports}
Our findings add controlled evidence to the informal practitioner remark relayed by InfoQ, in which Bryo AI describes Jev as considerably cheaper than Gemini while rating Gemini as slightly more accurate on email classification~\cite{infoq}. The remark does not state the Gemini version, the dataset, or the protocol, so it is not directly comparable with our run. In our experiment, Jev records higher accuracy than Gemini 3.5 Flash-Lite (79.67\% versus 76.20\%), and the cost ratio is approximately 2.2 (USD 0.032481 versus USD 0.015103 per 1000 decisions), which is far smaller than the ten- to twenty-fold difference described in that remark. These differences may reflect the model version, the dataset, or the prompt, and they illustrate why single anecdotal comparisons should be replaced by repeated, documented evaluations.

\subsection{Complementary Roles for Jev and LLMs}
We do not interpret the results as evidence that LLMs are ineffective. LLMs can produce explanations that help analysts understand why a message appears suspicious, and they can support broader reasoning tasks that extend beyond a fixed binary label~\cite{lee,kuikel}. Jev's typed output is useful when a workflow needs a structured decision and confidence value, but it does not provide the same free-form explanation. The systems therefore have complementary strengths.

One practical direction is a cascade architecture. A first-stage decision model can handle routine, clearly classified messages, while a second-stage LLM can review uncertain cases or produce explanations for human analysts. This study does not implement or measure such a cascade. We identify it as a hypothesis for future evaluation, in which researchers can vary the escalation threshold and measure the resulting accuracy, latency, cost, and analyst workload.

\subsection{Implications for Evaluation Practice}
The results show why phishing benchmarks should report more than accuracy. Precision and recall expose different error types, latency and API cost describe the operational effort, token usage helps explain request consumption, and calibration indicates whether confidence can support triage decisions. Future comparisons should report these dimensions together with raw predictions, confusion matrices, repeated trials, and enough implementation detail to support independent recalculation.

\section{Threats to Validity}\label{sec:threats}

We separate limitations that are documented in the experiment from risks that we have not tested but consider relevant to interpreting the results.

\subsection{Established Limitations}
We use one Kaggle dataset containing 1500 labelled emails, and the models receive a fixed set of metadata fields rather than full message bodies. The corpus supports a controlled comparison, but it does not represent the full diversity of enterprise email, languages, industries, or evolving attacks such as QR-code phishing and wire-transfer BEC. Because precision and accuracy depend on class prevalence, the near-balanced class distribution also limits how far the absolute scores can be compared with results on other corpora. The results therefore describe this fixed corpus and do not establish generalization to other email populations.

We report one evaluation pass per model and do not calculate run-to-run variance. Provider-side sampling, transient service load, and network variability can affect LLM outputs and latency. Repeated trials are necessary to determine whether the observed ranking remains stable, particularly for pairs with small differences such as the ECE of Jev and NVIDIA Nemotron 3 Ultra.

We use one equivalent prompt template across the three LLMs. This approach supports consistency but does not include model-specific prompt optimization, few-shot examples, or fine-tuning. The reported LLM scores are therefore baseline results under the selected prompt rather than each model's best possible performance. We also measure API calls over a standard internet connection rather than in an isolated inference environment, so the latency values combine provider response and network effects.

The LLM confidence values are verbalized in the generated text rather than derived from token probabilities, and we convert them to phishing probabilities with fixed rules, including a default of 0.5 when no confidence can be parsed. Jev's confidence is produced by a different mechanism, so the calibration comparison should be read as a comparison of reported confidence and not of a common underlying quantity. Different parsing rules or a different binning strategy could change the calibration scores, which is why the scripts and per-email outputs are made available.

\subsection{Potential Risks}
We interpret generated LLM responses as binary decisions with confidence values, whereas Jev returns a typed decision directly. If the interpretation rule treats ambiguous responses differently across providers, the LLM scores could be affected in ways that we cannot quantify from the current results.

The dataset is a public Kaggle listing from an individual contributor, and we have no documentation of how its labels were assigned. Any labelling inconsistencies would affect all four systems, but they could change the apparent ranking in ways that we cannot estimate.

Provider APIs, model versions, and prices change over time. The reported latency and costs apply to the evaluated versions and pricing conditions, and they should not be treated as permanent properties of the model families. A new evaluation should record the exact model identifiers, provider endpoint versions, pricing date, and request settings.

\section{Conclusion and Future Work}\label{sec:conclusion}
We compare Jev 1.13 with Gemini 3.5 Flash-Lite, NVIDIA Nemotron 3 Ultra, and Meta Llama 4 Maverick on a fixed dataset of 1500 labelled phishing and legitimate emails. We evaluate classification performance, latency, token consumption, API cost, and probability calibration. In this experiment, Jev records the highest accuracy (79.67\%), precision (71.43\%), recall (97.97\%), and F1 score (82.62\%). It also completes the dataset in 173.88 seconds, consumes 597,895 reported tokens, costs USD 0.015103 per 1000 decisions, and records the lowest Brier score (0.1698) and ECE (0.1594). These results support further investigation of Jev as an efficient option for high-volume, well-defined phishing triage. They do not establish universal superiority, production readiness, or replacement of LLMs. The LLMs retain value when analysts need generated explanations or broader reasoning, while Jev's typed decision interface suits structured classification workflows. We therefore view the strongest deployment hypothesis as a complementary design and not as a direct substitution.

Four directions warrant future investigation. First, developing a hybrid framework where Jev handles routine cases and escalates low-confidence predictions to an LLM. Second, evaluating larger, multilingual datasets incorporating complete messages and emerging phishing techniques. Third, conducting repeated trials and model-specific prompt optimisation to assess variability and strengthen baselines. Finally, evaluating production-scale deployment over time to examine changes in classification performance, latency, cost, and calibration as LLM models and pricing evolve.\\

\textbf{Replication Package.} We provide a replication package \cite{malik2026replication} containing the scripts and data necessary to reproduce, validate, and extend the results presented in this paper.

\enlargethispage{4\baselineskip}
{\footnotesize
\bibliographystyle{IEEEtran}
\bibliography{references}
}

\end{document}